\documentclass{article} % For LaTeX2e
\usepackage{nips15submit_e,times}
\usepackage{hyperref,amsmath,amsfonts,amssymb,multirow,graphicx,subfigure}
\usepackage{url}

\title{Information Relaxations and Dynamic Zero-Sum Games }

\author{
Martin B.~Haugh \\ %\thanks{ Use footnote for providing further information
Department of IE \& OR\\
Columbia University\\
New York, NY 10027 \\
\texttt{mh2078@columbia.edu} \\
\And
Chun Wang \\
Department of IE \& OR \\
Columbia University\\
New York, NY 10027 \\
\texttt{cw2519@columbia.edu} \\
}

\nipsfinalcopy % Uncomment for camera-ready version

\def\cA{\mathcal{A}}
\def\cB{\mathcal{B}}
\def\cF{\mathcal{F}}

\def\cL{\mathcal{L}}

\def\E{\mathbb{E}}

\def\b1{\mathbf{1}}

\def\hJ{\hat{J}}

\def\hz{\hat{z}}

\def\hbv{\hat{\mathbf{v}}}

\DeclareMathAlphabet{\mathsfsl}{OT1}{cmss}{m}{sl}
\renewcommand{\v}[1]{\boldsymbol{#1}}
\newcommand{\sss}[1]{\scriptscriptstyle{#1}}

\newcommand{\field}[2]{\mathbb{#1}^{#2}}

\newcommand{\evg}[1]{\mathbb{E}_{\sss{\mathcal{G}_{#1}}}}

\newtheorem{prop}{Proposition}

\begin{document}
% see sandholms science article. adversarial (cybersecurity, diseases), POMDPs in MDPs, poker-headsup

\maketitle
\begin{abstract}
Dynamic zero-sum games are an important class of problems with applications ranging from evasion-pursuit and heads-up poker to certain adversarial versions of control problems such as multi-armed bandit and multiclass queuing problems. These games are generally very difficult to solve even when one player's strategy is fixed, and so constructing and evaluating good sub-optimal policies for each player is an important practical problem. In this paper, we propose the use of information relaxations to construct dual lower and upper bounds on the optimal value of the game. We note that the information relaxation approach, which has been developed and applied successfully to many large-scale dynamic programming problems, applies immediately to zero-sum game problems. We provide some simple numerical examples and identify interesting issues and complications that arise in the context of zero-sum games.
\end{abstract}

\section{Introduction} \label{sec:intro}

Zero-sum games (hereafter ZSG's) have played a central role in the development of game theory with seminal contributions from Borel~\cite{Borel}, von Neumann~\cite{vonN1928} and Shapley~\cite{Shapley1953}. The latter considered discounted dynamic ZSG's and identified conditions under which they have a unique equilibrium value while Patek and Bertsekas~\cite{Pat-Bert} extended these results to more general stochastic shortest path (SSP) games.
ZSG's have many applications including pursuit-evasion / hunter-prey problems, heads-up poker and many so-called parlor games. Many other important problems, however, can be also modeled as ZSG's. These problems generally have an adversarial feature and applications can be found in many fields ranging from business to cyber-security and disease control. In addition, risk-averse control problems can also be cast as ZSG's by assuming that nature plays an adversarial role and selects probability distributions, payoffs or some other feature with a view to thwarting the decision-maker. This is a common approach in the multi-armed bandit\footnote{See Bubeck and Cesa-Bianchi~\cite{BubeckCesa} for an overview.} literature, for example, where the goal is to tradeoff the benefits of exploitation and exploration.

While standard dynamic programming (DP) methods such as value and policy iteration have been adapted to solve ZSG's, these methods quickly become computationally intractable as the game size grows. Indeed many important games of interest are too large to be solved exactly and require approximate solution approaches. Texas hold'em heads-up poker, for example, is a popular game and only very recently has the ``limit'' version of this game been effectively solved; see Bowling et al.~\cite{Bowling2015} as well as Sandholm~\cite{Sandholm2010} for an overview of approaches to tackling incomplete information ZSG's. Many DP problems are also intractable and this of course has led to the development of the approximate DP literature. Adversarial versions of these problems will clearly then also be intractable.

It is therefore important that we be able to construct good sub-optimal policy pairs (one for each player) for large-scale ZSG's. It is generally straightforward to simulate a policy pair and therefore obtain an unbiased estimate for the game value of this pair. But how far is this value from the (optimal) game value? While we can't answer this question unless we can solve for the game value itself, we can obtain bounds. In particular, if we fix one player's strategy we can then solve for the other player's best response and the resulting value will be a bound\footnote{Depending on which player's strategy is fixed, the bound will be a lower or upper bound on the game value.} on the game value. Unfortunately, solving for the best response is a DP problem that will itself often be intractable.

The goal of this paper is to demonstrate how tight bounds on these best-response problems can be found so that we can still obtain lower and upper bounds on the optimal game value. We do this via the recently developed information relaxation approach to obtaining dual bounds for DP problems. Moreover, strong duality (see part (c) of Prop. \ref{prop:bss}) tells us that if we start off with a policy pair that is close to optimal, then it should be possible to obtain tight lower and upper bounds on the optimal game value. Such bounds would then ``certify'' just how good the sub-optimal policy pair is.

The information relaxation approach was introduced independently by Brown, Smith and Sun~\cite{BrownEtAl2010} (hereafter BSS) and Rogers~\cite{Rogers2007} who in turn were partly motivated by earlier work ~\cite{rogers2002monte,haughkogan} on the pricing of high-dimensional American options. Over the past several years there have been many successful applications of this methodology; see for example ~\cite{BrownSmith2011,LaiEtAl,MoallemiSaglam,BrownSmith2013,HaughIyenWang,bender2011dual,schoenmakers2010pure,BrownHaugh2014,Hambly}. This paper was also motivated in part by Beveridge and Joshi~\cite{BeveridgeJoshi2011} who used the aforementioned work on American options to compute dual bounds for zero-sum optimal stopping games. This paper explains how dual bounds can be computed for general ZSG's as long as these games have an optimal value.

The remainder of this paper is organized as follows. We formulate the ZSG problem and review the results of Shapley~\cite{Shapley1953} in Section \ref{sec:ZSG-Form} and then review the information relaxation approach for constructing dual bounds in Section \ref{sec:InfoRelax-Review}. We  provide two examples in Section \ref{sec:egs}: a simple 2-period matrix game and an industrial-waste inspection game. We briefly discuss some specific issues and challenges that arise in the context of ZSG's in Section \ref{Sec:adv} and we conclude in Section \ref{sec:conc}.

\section{Dynamic zero-sum games}
\label{sec:ZSG-Form}

We now provide a brief overview of ZSG's and we mainly use the same notation and simplifying assumptions (finite action sets and state spaces, common information, simultaneous moves etc.) of Section 7.2 of Bertsekas and Tsitsiklis~\cite{Bert-Tsit}. We note, however, that there is no problem relaxing these assumptions. We essentially only require that the ZSG has a value in order to apply the information relaxation approach to compute dual bounds.

There are two players: we refer to player \emph{A} as the {\em maximizer} and player \emph{B} as the {\em minimizer}. At each time $t$ and in each state $i$, \emph{A} and \emph{B} choose actions $u$ and $v$ from finite constraint sets, $U(i)$ and $V(i)$, respectively. Transition probabilities are of the form $p_{ij}(u,v)$ and one-stage costs are given by $g(i,u,v,j)$. The players use randomized strategies to select $u$ and $v$ in each period. In particular, in state $i$ player \emph{A} chooses a probability distribution $y = \{ y_u \, | \, u \in U(i)\}$ over the set $U(i)$ while \emph{B} chooses a probability distribution $z = \{ z_v \, | \, v \in V(i)\}$. The system therefore moves from state $i$ to state $j$ with probability
\[
\sum_{u\in U(i)} \sum_{v\in V(i)}  y_u z_v p_{ij}(u,v)
\]
and the stage cost is
\[
G(i,y,z) = \sum_{u\in U(i)} \sum_{v\in V(i)}  y_u z_v  \sum_{j=1}^n p_{ij}(u,v)g(i,u,v,j).
\]
A policy $\pi_A = \{\mu_0, \mu_1, \ldots \}$ for player \emph{A} is a sequence of functions, $\mu_k$, so that \emph{A} selects his time $k$, state $i$ action via the probability distribution, $\mu_k(i)$ which is defined over $U(i)$. Similarly we use $\pi_B = \{ \nu_0, \nu_1, \ldots\}$ to denote a policy for player \emph{B}. Stationary policies have the form $\{\mu, \mu, \ldots \}$ and $\{\nu, \nu, \ldots \}$ and we simply refer to them as $\mu$ and $\nu$, respectively. The cost-to-go for policies $\pi_A$ and $\pi_B$ starting from state $i$ is
\[
J^{\pi_A, \pi_B}(i) = \sum_{k=1}^\infty \alpha^k \E\left[G(i_k,\mu_k(i_k),\nu_k(i_k))  \, | \, i_0 = i\right]
\]
where $\alpha \in (0,1)$ is the discount factor. We use $J^{\mu,\nu}(i)$ to denote the cost-to-go corresponding to the pair of stationary policies $(\mu,\nu)$. We can then define the min-max and max-min costs as
\begin{eqnarray*}
\underline{J}(i) &=& \min_{\pi_B} \max_{\pi_A} J^{\pi_A, \pi_B}(i) \\
\bar{J}(i) &=& \max_{\pi_A} \min_{\pi_B} J^{\pi_A, \pi_B}(i).
\end{eqnarray*}
We note that $\underline{J}(i)$  and $\bar{J}(i)$ are the optimal game values corresponding to the different orders in which \emph{A} and \emph{B} must choose their policies. The question then arises as to whether $\underline{J}(i)$ and $\bar{J}(i)$ are equal in which case we could define this common value to be the equilibrium value of the game. The key result of Shapley~\cite{Shapley1953} is that this is indeed the case. In particular, there exists a pair of stationary policies $(\mu^*,\nu^*)$ such that
\begin{equation} \label{eq:Shap}
J^{\mu^*,\nu^*}(i) = \max_{\mu} J^{\mu,\nu^*}(i) = \min_{\nu} J^{\mu^*,\nu}(i) = \underline{J}(i) = \bar{J}(i) =: J^*(i)
\end{equation}
where we use $J^*(i)$ to denote the equilibrium value of the game. Given (\ref{eq:Shap}) it is not surprising that the various solution approaches to dynamic programming problems, e.g. value iteration and policy iteration, have natural analogs for ZSG's. Details are given in \cite{Bert-Tsit} and for more general stochastic-shortest path games in \cite{Pat-Bert}. Since finite horizon games can formulated as infinite horizon games by including time $t$ as a state variable, Shapley's result also applies to these games resulting in optimal policies that are now time-dependent.

\subsection{Computing bounds on the game value}
\label{sec:BoundShapley}

Let $(\hat{\mu}, \hat{\nu})$ be any pair of stationary policies. Then it follows immediately from (\ref{eq:Shap}) that
\[
J^{\hat{\mu}}(i) \leq J^*(i) \leq J^{\hat{\nu}}(i)
\]
where $J^{\hat{\mu}}(i)$ is \emph{B}'s best response to $\hat{\mu}$ and $J^{\hat{\nu}}(i)$ is \emph{A}'s best response to $\hat{\nu}$. In theory both $J^{\hat{\mu}}(i)$ and $J^{\hat{\nu}}(i)$ can be found using standard DP methods such as value or policy iteration. For large scale games, however, computing $J^{\hat{\mu}}(i)$ and $J^{\hat{\nu}}(i)$ (as well as $J^*(i)$) may be intractable. For games of practical interest, this is often the case and so the best we can hope to do is to find good policies, $(\hat{\mu}, \hat{\nu})$, and bounds on $J^{\hat{\mu}}(i)$ and $J^{\hat{\nu}}(i)$. In particular, we would like to find a good lower bound, $J_{low}^{\hat{\mu}}(i)$, for $J^{\hat{\mu}}(i)$ and a good upper bound, $J_{up}^{\hat{\nu}}(i)$, for $J^{\hat{\nu}}(i)$. In that case we will have
\begin{equation} \label{eq:ShapBounds}
J_{low}^{\hat{\mu}}(i) \leq J^{\hat{\mu}}(i) \leq J^*(i) \leq J^{\hat{\nu}}(i) \leq J_{up}^{\hat{\nu}}(i)
\end{equation}
and it follows that $[J_{low}^{\hat{\mu}}(i), \, J_{up}^{\hat{\nu}}(i)]$ is an interval containing the true value of the game, $J^*(i)$. We note that finding a good lower bound for $J^{\hat{\mu}}(i)$ is not straightforward since $J^{\hat{\mu}}(i)$ is the solution of a DP where the goal is to {\em minimize} costs. Any feasible policy for this DP therefore yields an upper bound for $J^{\hat{\mu}}(i)$ so that the inequality goes in the wrong direction. A similar observation applies to finding an upper bound for $J^{\hat{\nu}}(i)$. In the next section we describe how the desired bounds, $J_{low}^{\hat{\mu}}(i)$ and $J_{up}^{\hat{\nu}}(i)$, can be computed using the concept of information relaxations and dual penalties.

\section{Information relaxations for dynamic programming problems}
\label{sec:InfoRelax-Review}

The information relaxation approach was developed independently by BSS~\cite{BrownEtAl2010} and Rogers~\cite{Rogers2007} but the former is more general and we follow\footnote{Our overview of BSS follows Brown and Haugh~\cite{BrownHaugh2014}.} their approach here. In particular we will consider a DP problem where the objective is to minimize total expected costs. This is the {\em primal} DP problem and it is exactly this problem\footnote{By considering a DP problem where the goal is to maximize total expected gains we can use the information relaxation approach developed in this section to obtain an upper bound on $J^{\hat{\nu}}(i)$. The approach is identical so we will consider only the construction of the dual lower bound here.} that we face in trying to obtain a lower bound on $J^{\hat{\mu}}(i)$ as described above.
In order to keep the notation manageable, however, we will drop all references to $\hat{\mu}$ as well as any reference to the players. We will begin with finite horizon problems after which we will describe how infinite horizon problems can be handled. It will also be convenient to introduce some alternative notation to that developed in Section \ref{sec:ZSG-Form}. In particular, we will now use $x \in X$ to denote a generic state and we will assume a state transition function
\begin{equation} \label{eq:BSS-state-dyn}
x_{t+1} = s(x_t,v_t,w_{t+1})
\end{equation}
where $x_t$ is the time $t$ state, $v_t$ is the action chosen at time $t$ and $w_{t+1}$ is a random variable. Without loss of generality we can take the $w_t$'s to be IID $U(0,1)$ random\footnote{To connect with the formulation of Section \ref{sec:ZSG-Form}, we can take $w_t$ to be the random variable that is used to generate the state transition in accordance with $p_{x_t,x_{t+1}}(\hat{\mu}(x_t),v_t)$ where we are keeping player A's policy fixed at $\hat{\mu}$. While these probabilities depend on the action, $v_t$, we can still take the $w_t$'s to be IID and have the state dependence handled via the transition function, $s$.} variables.

The information relaxation approach considers policies that use more information than is available to primal feasible policies. Specifically, we consider policies that can use advance information about realizations of the uncertain sequence, $\{w_t\}_{t\geq 1}$; such policies are generally not feasible to the primal DP. By finding the best such policies, we can obtain lower bounds on the optimal value of the primal DP.
To make this idea concrete, we let $\mathbb{F}$ denote the underlying filtration associated with $\{w_t\}_{t\geq 1}$, i.e., $\mathbb{F}:=\{\mathcal{F}_t\}_{t\geq 0}$, where $\mathcal{F}_t$ denotes the $\sigma$-algebra generated by $(w_1,\ldots,w_t)$ at time $t$. We will describe additional information structures via another filtration $\mathbb{G}:=\{\mathcal{G}_t\}$. We say $\mathbb{G}$ is a \emph{relaxation of $\mathbb{F}$} if $\mathcal{F}_t\subseteq\mathcal{G}_t$ for all $t$, i.e., under $\mathbb{G}$, at least as much (and perhaps more) is known about $\{w_t\}_{t\geq 1}$ at all times $t$. We let $\evg{t}$ denote expectations conditional on $\mathcal{G}_t$. We will refer to realizations of $\{w_t\}_{t\geq 1}$ as \emph{scenarios} or \emph{dual paths}.
A policy is adapted to a filtration $\mathbb{F}$ (or $\mathbb{F}$-adapted) if the actions taken by the policy in every period $t$ are measurable with respect to $\mathcal{F}_t$. We let $\mathcal{U}_{\mathbb{F}}$ denote the set of feasible, $\mathbb{F}$-adapted policies. If $\mathbb{G}$ is a relaxation of $\mathbb{F}$, then $\mathcal{U}_{\mathbb{F}}\subseteq\mathcal{U}_{\mathbb{G}}$: a policy that is $\mathbb{F}$-adapted is also $\mathbb{G}$-adapted.

Given a finite horizon, $T$, the expected time $t=0$ cost for any $\nu\in\mathcal{U}_{\mathbb{F}}$ can be written as
\begin{eqnarray*}
J_{\sss{0}}^{\nu}(x_{\sss{0}}) = \mathbb{E}\big[\sideset{}{_{t=0}^{\sss{T-1}}}\sum G_t(x_t,\nu_t(x_t))\big].
\end{eqnarray*}
We let $J_t(x_t)$ denote the optimal cost at time $t$ from state $x_t$ and we also define $\v{v}:=(v_0,\ldots,v_{T-1})$ and $\v{w}:=(w_1,\ldots,w_T)$. BSS then consider penalty functions $z(\v{v},\v{w}):=\sum_{t=0}^{T-1} z_t(\v{v},\v{w})$ that depend on actions and uncertainties. They call a penalty function $z$ \emph{dual feasible} if $\mathbb{E}[z(\nu)]\leq 0$ for all $\nu\in\mathcal{U}_{\mathbb{F}}$. The following result recaps some of the main results from BSS.

\begin{prop}\label{prop:bss} {\bfseries (BSS 2010)}
Let $\mathbb{G}$ be any information relaxation of $\mathbb{F}$.
\begin{itemize}
\item[(a)] \emph{Weak duality:} For any dual feasible penalty $z$,
\begin{eqnarray}\label{eq:bss_weak_duality}
J_{\sss{0}}(x_{\sss{0}}) &\geq& \inf\limits_{\nu\in\mathcal{U}_{\mathbb{G}}} \mathbb{E}\big[\sideset{}{_{t=0}^{\sss{T-1}}}\sum G_t(x_t,\nu_t(x_t)) + z_t(\nu)\big].
\end{eqnarray}

\item[(b)] \emph{Dual feasible penalties:} Let $h_1,\ldots,h_T$ be any sequence of measurable functions $h_t:X\rightarrow\field{R}{}$. Then the penalty given by $z_t(\v{v},\v{w})=\mathbb{E}[h_{t+1}(s_{t+1}(x_t,v_t,w_{t+1}))]-\evg{t}[h_{t+1}(s_{t+1}(x_t,v_t,w_{t+1}))]$ is dual feasible.

\item[(c)] \emph{Strong duality:} When $h_t=J_t$, inequality (\ref{eq:bss_weak_duality}) holds with equality.
\end{itemize}
\end{prop}

Part (a) of Prop. \ref{prop:bss} shows that we can get lower bounds with any information relaxation and any dual feasible penalty. For example, when $\mathbb{G}$ is the perfect information (PI) relaxation where the entire scenario $\v{w}$ is revealed at $t=0$, we can simply simulate scenarios and select actions ``path-wise'' that minimize the sum of costs plus penalties. The expected value of the optimal costs (plus penalties) then provides a lower bound on the optimal value $V_{\sss{0}}(x_{\sss{0}})$. More specifically, assuming a PI relaxation we can compute an unbiased lower bound on $J_{\sss{0}}(x_{\sss{0}})$ by simulating $N$ scenarios, $\v{w}^{n} = (w_1^{n},\ldots,w_T^{n})$ for $n=1, \ldots , N$, and then solving for
\begin{equation} \label{eq:dual}
\underline{J}_0^{(n)} := \inf\limits_{v_0,\ldots,v_{T-1}} \,   \sideset{}{_{t=0}^{\sss{T-1}}}\sum G_t(x_t,v_t) + z_t(\v{v},\v{w}^{n}).
\end{equation}
where the $x_t$'s satisfy (\ref{eq:BSS-state-dyn}) with $w_t^{(n)}$ replacing $w_t$. (With a PI relaxation the ``inf'' in (\ref{eq:bss_weak_duality}) can be moved inside the expectation which then yields a ``dual'' or ''inner'' problem of the form (\ref{eq:dual}).) An unbiased lower bound for $J_0$ is then given by $\sum_{n=1}^N \underline{J}_0^{(n)} / N$.

Part (b) of Prop. \ref{prop:bss} is useful because it provides a method for constructing dual feasible penalties: we can obtain a dual feasible penalty by adding up differences in conditional expectations of a sequence of any ``generating functions'' $h_1,\ldots,h_T$. Finally, part (c) of Prop. \ref{prop:bss} shows that this approach, in theory, provides tight bounds: by taking the generating functions to be the optimal cost functions, the optimal value from the relaxed problem equals the optimal cost of the primal DP. Of course, when applying the method, we would typically not know $J_t$, but we can take the generating functions to be approximate value functions, $\hat{J}_t$, and, by parts (a) and (b) of Prop. \ref{prop:bss}, we will nonetheless obtain a lower bound on the optimal cost. Moreover, the closer $\hat{J}_t$ is to $J_t$ the better we expect the lower bound to be. As mentioned in Section \ref{sec:intro}, there have now been many successful applications of this methodology in the DP literature. We note that BSS also showed that for any given penalty, $z_t(\v{v},\v{w})$, information relaxations of $\mathbb{F}$ that reveal less information yield tighter dual bounds.

\subsubsection*{Infinite horizon problems}
%\label{sec:InfHor}

As stated earlier, Shapley~\cite{Shapley1953} solved the infinite horizon discounted ZSG problem (which includes the finite-horizon ZSG as a special case). While the information relaxation approach was originally developed for finite horizon DP's, Brown and Haugh~\cite{BrownHaugh2014} show how it can easily be extended to infinite horizon discounted problems. The industrial-waste inspection game below, however, is not an infinite horizon discounted ZSG but is instead a stochastic-shortest path (SSP) ZSG. An SSP ZSG is a non-discounted infinite horizon game that has a terminal absorbing state that may or may not be reached in a finite number of time periods. The theory of such games is more delicate and was treated\footnote{In his thesis Patek~\cite{Patek1997} considered the industrial-waste game and showed that it has an optimal game value} by Patek and Bertsekas~\cite{Pat-Bert}.

One potential difficulty that arises in constructing dual bounds for SSP ZSG's is in simulating a dual sample path, $(w_1,\ldots, w_T)$. The absorption time, $T$, is random and policy-dependent and so it's not clear how to simulate such a path. The weak-form approach of Rogers~\cite{Rogers2007} and Brown and Haugh~\cite{BrownHaugh2014} show how this problem can\footnote{Brown and Haugh~\cite{BrownHaugh2014} identify other approaches such as their truncated horizon or quasi-strong-form approaches that can often yield superior bounds.} be resolved. In particular, dual sample paths are simulated under a reference transition probability measure, $Q$, that is {\em action-independent}. We can use such a $Q$ to generate dual sample paths as long as all such paths terminate after a finite number of periods $Q$-almost surely. Appropriate Radon-Nikodym terms are then used to adjust the objective function in (\ref{eq:dual}) to account for this change of measure. It is also necessary to ensure certain absolute-continuity conditions are either satisfied
or handled appropriately. Space constraints prevent us from expanding further on these issues but see ~\cite{BrownHaugh2014} for further details.

\section{Some Examples}
\label{sec:egs}

We now consider two examples where we use the information relaxation methodology to construct dual bounds on the optimal game values. These are simply illustrative examples so in each case we were in fact able to solve for both the best responses, $\max_{\mu} J^{\mu,\hat{\nu}}(i)$ and $\min_{\nu} J^{\hat{\mu},\nu}(i)$, as well as the optimal game value. The dual bounds were constructed in each case using the value function, $J^{\hat{\mu},\hat{\nu}}(i)$, corresponding to some fixed policy pair $(\hat{\mu},\hat{\nu})$, to construct dual feasible penalties. We do note, however, that approximate value functions can be constructed in many different ways and that it may be advantageous to use different approximate value functions (and indeed different transition measures, $Q$) to construct the dual penalties required for each of the two dual bounds.

\subsection{A dynamic matrix game}
\label{sec:EG-2period}

We first consider a single-period two-person ZSG where the payoff is defined by an $m \times n$ matrix, $R$. In this game players \emph{A} and \emph{B} simultaneously select a row, $u$, and a column, $v$, respectively, after which \emph{B} pays \emph{A} the value $R_{u,v}$. A mixed policy, $y$, for player \emph{A} is an $m \times 1$ vector such that \emph{A} chooses the $u$-th row with probability $y_u$. Similarly, we let $z$ denote player \emph{B}'s mixed policy. The expected payoff to \emph{A} is then $y'R z$.

Consider now a $2$-period dynamic game defined by the following three matrix games:
\begin{equation}
\label{eq:FiniteGame-PayoffMatrix-R1R2R3}
R^{(1)} =\begin{bmatrix} 2 & 1 \\ 6 & 8 \end{bmatrix} \ \ \ \ \
R^{(2)} =\begin{bmatrix} 8 & 15 \\ 10 & 12 \\ \end{bmatrix} \ \ \ \ \
R^{(3)} =\begin{bmatrix} -8 & -10 \\ 3 & -11 \\ \end{bmatrix}. \nonumber
\\
\end{equation}
The state variable $x_{t} \in \{i \, | \, i=1,2,3\}$ determines the game $R^{(x_t)}$ that is played at time $t$, and  $G(x_{t},u_{t},v_{t})$ is then the payoff of that game when \emph{A} and \emph{B} choose the $u_{t}^{th}$ row and $v_{t}^{th}$ column, respectively. The game begins at $t=0$ with \emph{A} and \emph{B} playing $R^{(1)}$ so the initial state is $x_{0} = 1$. The game to be played at $t=1$ is determined by \emph{A} and \emph{B}'s actions and a transition probability, $p_{x_{t},x_{t+1}}(u_{t},v_{t})$. These transition probabilities are determined by the $(u_{t}, \, v_{t})$ element in the matrices $P_{x_{t},x_{t+1}}$ which are defined as:
\begin{equation}
\label{eq:FiniteGame-TransitionMatrix}
P_{1,2} =\begin{bmatrix} 0.7 & 0.55 \\ 0.4 & 0.5 \end{bmatrix} \text{ and }
P_{1,3} =\begin{bmatrix} 0.3 & 0.45 \\ 0.6 & 0.5 \end{bmatrix} . \nonumber
\end{equation}
For example, if \emph{A} chooses the second row  and \emph{B} chooses the first column when playing $R^{(1)}$ at time $0$, then \emph{B} pays \emph{A} $6$ units and  at time $t=1$ they will play game $R^{(2)}$ with probability $0.4$ and game $R^{(3)}$ with probability $0.6$.
%With $\omega_{1} \sim U(0,1)$ the state transition can be written as
%\begin{equation}
%x_1 = f_{0}(x_{0},a_{0},b_{0},\omega_{1}) = 2\cdot\b1_{\{\omega_{1} \le p_{1,2}(a_{0},b_{0})\}} + 3\cdot(1-\b1_{\{\omega_{1} \le p_{1,2}(a_{0},b_{0})\}})
%\end{equation}
%where $\b1_{\{\cdot\}}$ denotes the indicator function.
The optimal policies and value functions for this 2-period ZSG are easily calculated using standard techniques and are given in Table \ref{table:MatrixGameOptimalTable}.

\begin{table}
\begin{centering}
\caption{Optimal policies and game values}
\label{table:MatrixGameOptimalTable}
\begin{tabular}{c c c c r }
\hline
$t$ & $x_{t}$ & $u_{t}^{*}$ & $v_{t}^{*}$ & $J_{t}(x_{t})$ \\
\hline
$0$ & $1$ & $[0.5 \quad 0.5]'$ & $[0.75 \quad 0.25]'$ & $5$ \\
$1$ & $2$ & $[0 \quad 1]'$ & $[1 \quad 0]'$ & $10$ \\
$1$ & $3$ & $[1 \quad 0]'$ & $[0 \quad 1]'$ & $-10$ \\
\hline
\end{tabular}
\par\end{centering}
\end{table}

We can compute an upper bound for the fair value of the game, $J_0(1)$, by fixing \emph{B}'s policy, $\hat{\nu}$, and then solving for \emph{A}'s best response. If we take $\hat{\nu} = \left\{\hat{\nu}_{0}(x_0=1) = [0.6 \quad 0.4]',  \, \hat{\nu}_{1} = v_{1}^{*}(x_{1})\right\}$, then  the solution to \emph{A}'s best response DP problem is given by:
\begin{eqnarray}
J_{1}^{\hat{\nu}}(x_{1})&=& J_{1}(x_{1}) \label{eq:FiniteGame-suboptimal-J-t1} \\
J_{0}^{\hat{\nu}}(x_{0}) &=& \max_{u\in{1,2}} \bigg\{R \, \hat{\nu}_{0} = \begin{bmatrix} 4.4 \\ 5.6 \end{bmatrix}  \bigg\} =5.6   \label{eq:FiniteGame-suboptimal-J-t0}
\end{eqnarray}
where
\[
R := R^{(1)}+J_{1}(2)P_{1,2}+J_{1}(3)P_{1,3} = \begin{bmatrix} 6 & 2 \\ 4 & 8 \end{bmatrix}.
\]
We note that $5.6 = J_{0}^{\hat{\nu}}(x_{0})$ is of course an upper bound on the value of game, $J_{0}(x_{0})=5$.
For more complex games we would not be able to compute $J_{0}^{\hat{\nu}}(x_{0})$ but we can instead use the information relaxation approach to compute an upper bound, $J_{up}^{\hat{\nu}}(x_0)$, on $J_{0}^{\hat{\nu}}(x_{0})$. Suppose we use a PI relaxation and construct the penalty $\hz = \hJ_{1}^{\hat{\nu}}(x_{1}) - \E\big[\hJ_{1}^{\hat{\nu}}(x_{1})\big|\cF_0\big]$ where $\hJ_{1}^{\hat{\nu}}(x_{1})$ is an approximate value function for player \emph{A}'s DP. For illustrative\footnote{This is the game value at time $t=1$ if \emph{A} always chooses first row and \emph{B} always chooses first column when playing $R^{(x_{1})}$.} purposes, we take $\hJ_{1}(x_{1}; \hbv) = 8\cdot\b1_{\{x_{1}=2\}} - 8\cdot\b1_{\{x_{1}=3\}}$.
We estimated the dual bound by simulating 10,000 values of $w_{1}$, solving the deterministic dual inner problem (\ref{eq:dual}) for each value, and then averaging the results. This yielded an estimated upper bound $J_{up}^{\hat{\nu}}(x_0; \hz(\hat{\nu})) = 5.82$ with a standard error of $0.02$.

As an aside, we note that if we instead used the dual optimal penalty $z^{*} = J_{1}^{\hat{\nu}}(x_{1}) - \E\big[J_{1}^{\hat{\nu}}(x_{1})\big|\cF_0\big]$, each dual inner problem (\ref{eq:dual}) yields an upper bound $J_{up}^{\hat{\nu}}(x_0; z^*(\hat{\nu})) = 5.6$, thereby demonstrating strong duality (for \emph{B}'s fixed policy, $\hat{\nu}$) as defined by part (c) of Prop. \ref{prop:bss}. Note also that if we fixed \emph{B}'s policy at the optimal $\nu^*$ and repeated the numerical calculations above, then we would obtain $J_{0}(x_{0}) = J_{0}^{\nu^*}(x_{0}) = J_{up}^{\nu^*}(x_0; z^*(\nu^*))=5$. Corresponding lower bounds for $J_{0}(x_{0})$ can be obtained analogously by fixing \emph{A}'s policy and solving or lower bounding player \emph{B}'s resulting DP.

\subsection{An industrial-waste inspection game}

We now consider a somewhat more practical game as discussed in Patek~\cite{Patek1997}. The two players are a manufacturer (player \emph{A}) who produces industrial waste that must be dumped every night, and an inspector (player \emph{B}) who wants to catch the manufacturer dumping. There is a finite number of geographically disparate sites where industrial waste can be dumped. \emph{A} must dump waste at one of these sites every night while avoiding detection by \emph{B}. In order to detect dumping activity on a given night, \emph{B} must inspect the same site where \emph{A} is dumping that night and even then, there is a nonzero probability of failing to catch \emph{A}. In particular, conditional on \emph{A} and \emph{B} selecting the same site, the probability of detection depends upon two considerations:
\begin{enumerate}
\item The closer the current dumping site is to the preceding dumping site, the greater the probability of detection. (This models the environment's limited ability to absorb waste.)
\item The closer the current inspection site is to the preceding inspection site, the greater the probability of detection. (This models the fact that the inspector needs less time to travel and so he has more time to look for dumping activity.)
\end{enumerate}
If \emph{B} detects \emph{A} dumping two nights in a row, then \emph{A} is put out of business. \emph{A}'s objective is to maximize its time in business, while \emph{B}'s objective is to minimize \emph{A}'s time in business. In deciding where to dump / inspect each night, we assume that both players know where the dumping and inspection occurred on the previous night. They also know whether or not \emph{A} was caught dumping the previous night. Patek~\cite{Patek1997} shows that this game has an optimal equilibrium value.

We can formulate this as a zero-sum SSP game as follows. Let $\cL = \{l_{1},...,l_{N}\}$ represent the $N$ sites where waste may be dumped.
We let $u_{t-1} \in \cL$ denote the site where \emph{A} dumped at time $t-1$ and we let $v_{t-1} \in \cL$ denote the site inspected by \emph{B} at that time. Let $c_{t}$ be a Boolean variable which is TRUE if \emph{A} was caught dumping at time $t-1$. The state vector $x_{t} = (u_{t-1},v_{t-1},c_{t})$ then describes the state of the system at time $t$. There are\footnote{Note that states $(u_{t-1},v_{t-1},c_{t}=1)$ are not possible unless $u_{t-1}=v_{t-1}$. Hence there are $2N^2 - N(N-1) = N^2+N$ possible states.} $N^2+N$ possible non-absorbing states and one absorbing state. Given the manufacturer \emph{A} has not yet been shut down at time $t$, \emph{A} chooses to dump at site $u_{t} \in \cA(x_{t}) = \cL$ and \emph{B} chooses to search at site $b_{t} \in \cB(x_{t})= \cL$. The probability that \emph{A} will then be detected on night $t$ is
\begin{equation}
\label{eq:Dumpgame-p}
p^{d}(x_{t},u_{t},v_{t}) = \begin{cases}\bar{p}^{d} + \frac{\underline{p}^{d} - \bar{p}^{d}}{(k_{1}+k_{2})\bar{d}}[k_{1}d(u_{t},u_{t-1})+k_{2}d(v_{t},v_{t-1})], & \text{if } u_{t} = v_{t}\\
0, & \text{otherwise}\end{cases}
\end{equation}
where $d(l_{i},l_{j})$ denotes the distance between sites $l_{i}$ and $l_{j}$, $\bar{d} = \max_{l_{i},l_{j} \in \cL}d(l_{i},l_{j})$ is the max distance between any two sites, $0 < \underline{p}^{d} < \bar{p}^{d} < 1$ are worst-case and ideal
%\footnote{$\bar{p}^{d}$ is possible only if there is some pair of sites, $l_{i}$ and $l_{j}$, for which $d(l_{i},l_{j})=0$.}
probabilities of detection, and $k_{1}$ and $k_{2}$ are positive weighs. If $c_{t}=$ FALSE, then the system transitions to state $(u_{t},v_{t},\text{TRUE})$ with probability $p^{d}(x_{t},u_{t},v_{t})$, otherwise the system transitions to $(u_{t},v_{t},\text{FALSE})$. If $c_{t}=$ TRUE, then the game transitions to the absorbing terminal state with probability $p^{d}(x_{t},u_{t},v_{t})$, otherwise the system transitions to $(u_{t},v_{t},\text{FALSE})$. The cost is 1 in each time period regardless of the controls applied\footnote{Any pure policy for the inspector \emph{B} is improper in this game because, knowing this policy, the manufacturer \emph{A} can always avoid being detected. On the other hand, a mixed policy which select each site with a positive probability will eventually result in \emph{A} getting caught, so there exists a proper policy. See Patek~\cite{Patek1997} for more on proper policies for SSP ZSG's.}.

We consider a case with $N=10$ sites that are located on a straight line with equal distances between successive sites and where $d(l_i,l_j) = |j-i|$. We assume parameter values of $\underline{p}^{d} = 0.5$, $\bar{p}^{d} = 0.95$, $k_{1} = 2$ and $k_{2}=1$.  We begin with suboptimal policies, $(\mu^0,\nu^0)$, where \emph{A} and \emph{B} select sites uniformly each night and we compute the game value, $J^{(\mu^0,\nu^0)}(x_0)$, for these policies. We then use the so-called naive policy iteration algorithm (see \cite{Patek1997} and \cite{Poll69}) to update these policies. We use $(\mu^k,\nu^k)$ to denote the policies resulting from $k$ rounds of policy iterations and the corresponding game values are denoted by $J^{(\mu^k,\nu^k)}(x_0)$. After each policy iteration, we also computed each player's best response to the other player's strategy.  For more complex games in general, these best responses cannot be calculated and so we therefore also computed dual bounds to these best response value functions. The dual bounds were computed using dual feasible penalties constructed according to part (b) of Prop. \ref{prop:bss} with $h(x_t)$ taken equal to $J^{(\mu^k,\nu^k)}(x_t)$.

\begin{figure}[h]
\begin{center}
\begin{centering}
%\framebox[4.0in]{$\;$}
\fbox{%\rule[-.5cm]{0cm}{4cm}
\includegraphics[width=0.55\linewidth]{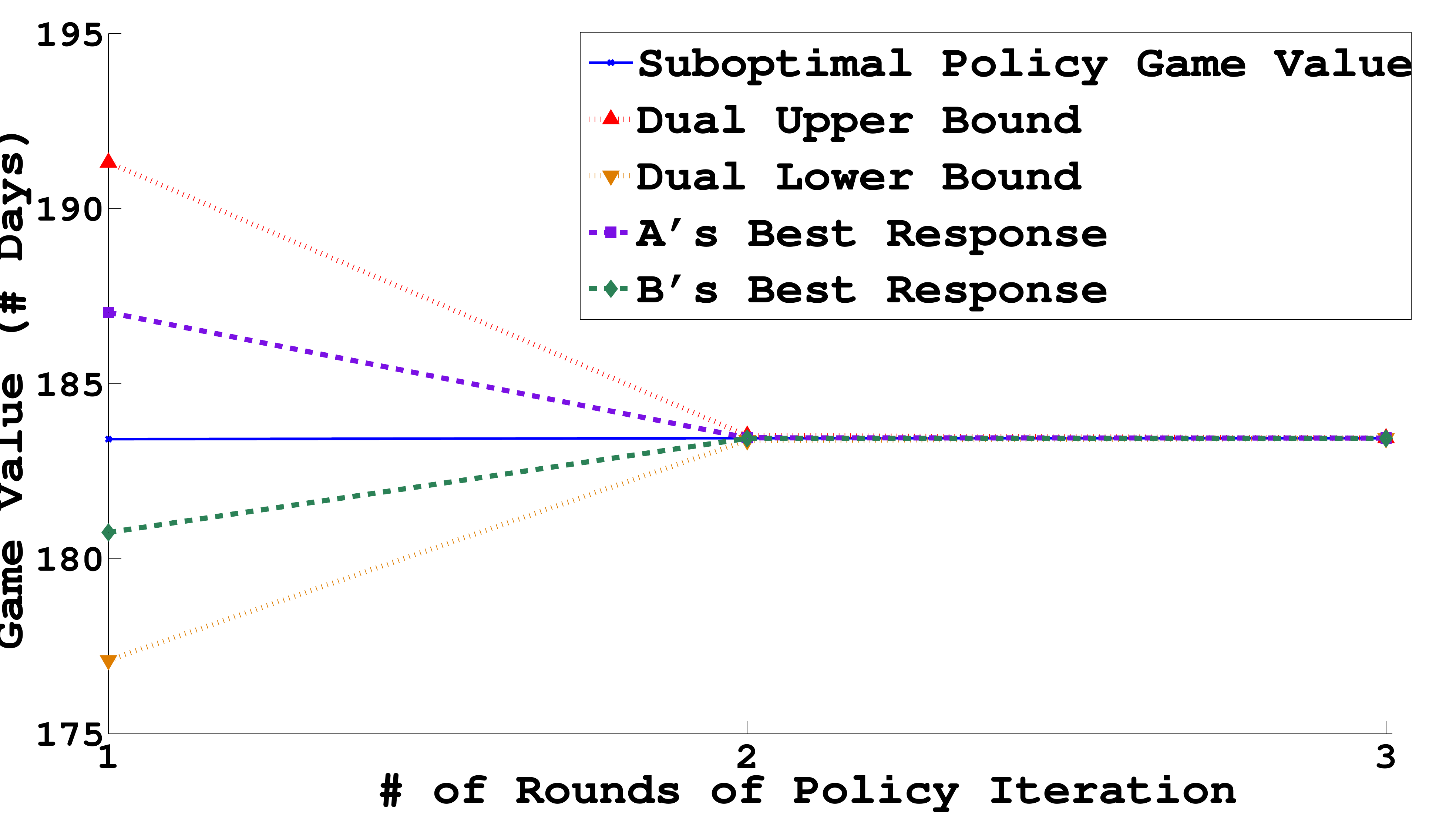} }
%\rule[-.5cm]{1cm}{0cm}}
%\includegraphics[width=5in]{N10.eps}
\caption{Dual bounds, best responses and game values for waste-inspection game}
\label{fig:Dumpgame}
\end{centering}
\end{center}
\end{figure}

The results are displayed\footnote{We did not plot the results for $k=0$ since in that case the bounds were very far apart -- the dual lower and upper bounds were 128.8 and 365.5, respectively -- and would disrupt the scale of the plot.} in Figure \ref{fig:Dumpgame} for the initial state $(l_{1},l_{1},\text{FALSE})$ and were computed using the weak-form dual approach of Rogers ~\cite{Rogers2007} and Brown and Haugh~\cite{BrownHaugh2014}. We used a $Q$ that was action-independent and that moved from any state (except the absorbing state) to any other state with equal probability, i.e. with probability $1/(N^2 + N)$. The terminal state was naturally also $Q$-absorbing.
Note that the dual bounds become very tight after just two rounds of policy iteration. This of course is consistent with strong duality, i.e. part (c) of Prop. \ref{prop:bss}, which suggests that dual penalties constructed from better approximations to the value function should yield tighter dual bounds.

\section{Adversarial games and games with imperfect information}
\label{Sec:adv}

There are several interesting features of the information relaxation approach that are even more pronounced for ZSG's. We briefly discuss some of these features now. Many of the most interesting ZSG's are imperfect information games where one or both players have private information. The best response of each player is then a partially observed Markov decision process (POMDP). It is well known (see \cite{Astrom}) that these problems can be converted into DP problems and so these games fit into our framework. That said, there is considerably more flexibility regarding the choice of information relaxation, $\mathbb{G}$, and we can expect this flexibility to greatly influence the tractability of the dual inner problems. Moreover, there is no reason to require that the dual lower and upper bounds are constructed using the same set of simulated paths, penalties, control variates (to reduce Monte-Carlo uncertainty) or transition measures, $Q$. Indeed, depending on the form of transition measures used to generate dual inner problems, it may be necessary to use different penalties for the dual lower and upper bounds.

We also note that adversarial versions of control problems can be considerably harder to handle. For example, Brown and Haugh~\cite {BrownHaugh2014} use Lagrangian relaxation methods to construct dual penalties and therefore obtain tight dual bounds for an intractable multiclass queuing problem. This approach might not work, however, for most adversarial versions of this problem where nature gets to choose the (presumably) state-dependent arrival rates. This introduces new challenges in finding suitable approximate value functions / dual penalties for these problems.

\section{Conclusions}
\label{sec:conc}

We have shown how the information relaxation approach for DP problems can be applied to intractable ZSG's where the best response DP problems are also intractable. This is an active area of research within the DP literature with many interesting questions still to be addressed and new challenges that arise in the context of large-scale ZSG's.

%\subsubsection*{Acknowledgments}
%
%Use unnumbered third level headings for the acknowledgments. All
%acknowledgments go at the end of the paper. Do not include
%acknowledgments in the anonymized submission, only in the
%final paper.

\newpage

\subsubsection*{References}
\renewcommand{\section}[2]{}%
\bibliographystyle{unsrt}
{\small
\bibliography{ZSG}
}
\end{document}